\documentclass[amsmath,amssymb,aps,reprint]{revtex4-1}

\usepackage{xcolor}

\usepackage{graphicx}
\usepackage{amssymb,amsmath,tabularx}

\begin{document}
\title{Stochastic social behavior coupled to COVID-19 dynamics leads to waves, plateaus and an endemic state
}
\author{ Alexei V. Tkachenko$^{2\dagger}$, Sergei Maslov$^{1, 4,5\dagger}$, Tong Wang$^{2,5}$,  Ahmed Elbanna$^{3}$, George N.~Wong$^{1}$, and Nigel Goldenfeld$^{1,5}$}

\affiliation{$^1$Center for Functional Nanomaterials, Brookhaven National Laboratory, Upton, NY 11973, USA\\
$^2$Department of Physics, University of Illinois at Urbana-Champaign, Urbana, IL 61801, USA\\
$^3$Department of Civil Engineering, University of Illinois at Urbana-Champaign, Urbana, IL 61801, USA\\
$^4$Department of Bioengineering, University of Illinois at Urbana-Champaign, Urbana, IL 61801, USA\\
$^5$Carl R. Woese Institute for Genomic Biology, University of Illinois at Urbana-Champaign, Urbana, IL 61801, USA}

\medskip
\date{\today}

\begin{abstract}
It is well recognized that population heterogeneity plays an important role in the spread of epidemics. While individual variations in social activity are often assumed to be persistent, i.e. constant in time, here we discuss the consequences of dynamic heterogeneity. 
By integrating the stochastic dynamics of social activity into traditional epidemiological models we demonstrate the emergence of a new long timescale governing the epidemic in broad agreement with empirical data.
Our model captures multiple features of real-life epidemics such as COVID-19, including prolonged plateaus and multiple waves, which are transiently suppressed due to the dynamic nature of social activity. 
%
%
The existence of the long timescale due to the interplay between epidemic and social dynamics provides a unifying picture of how a fast-paced epidemic typically will  transition to the endemic state. %
\end{abstract}
\maketitle

The COVID-19 pandemic has underscored the prominent role played by population heterogeneity in  epidemics.     On one hand, the observed transmission of infection  is  characterized by the phenomenon of super-spreading, in which a small fraction of individuals are responsible for a disproportionally large number of secondary infections \cite{LloydSmith2005,May_superspr_2005,Super_Kucharski,Vespigniani2020}. On the other hand, according to  multiple models, population heterogeneity is expected 
to suppress the herd immunity threshold (HIT) and reduce the final size of an epidemic 
\cite{pastor2015epidemic,bansal2007individual,Herd_Gomes,hetero_PNAS,neipel2020power,Herd_Science_2020}.
In the context of COVID-19, this observation led to a controversial suggestion that a strategy relying exclusively on quickly reaching herd immunity might be a viable alternative to government-imposed mitigation \cite{nature_herd2020}. However, the experience of locations that had embraced that strategy has exposed its flaws. While the first wave of infections in those locations never reached the scale of an unmitigated  epidemic predicted by classical homogeneous  models, it also failed to provide long-lasting protection against new waves \cite{buss2021three}. 

Another puzzling aspect of the  COVID-19 pandemic  is plateau-like  dynamics, where the  incidence rate stays at an approximately constant level for a prolonged time \cite{DIVOC,network_plat2020,Weitz2020_awareness}. These dramatic departures from predictions of both classical epidemiological models and their heterogeneous extensions have led to a greater  appreciation of the role played by human behavior in epidemic dynamics. In particular, one plausible mechanism that might be responsible for both suppression of the early waves and plateau-like dynamics is that individuals modify their behavior based on information about the current epidemiological situation \cite{PNAS2009_behavior,behavior_activity,Weitz2020_awareness}. Another possibility is that long plateaus might arise because of the underlying structure of social networks \cite{network_plat2020}.

Here we study epidemic dynamics accounting for random changes in levels of individual social activity. We demonstrate that this type of dynamic heterogeneity, even without  knowledge-based adaptation of human behavior (e.g. in response to epidemic-related news) \cite{PNAS2009_behavior,behavior_activity,Weitz2020_awareness}, leads to a substantial revision of the epidemic progression, consistent with the  empirical data for the COVID-19 pandemic.  In a recent study \cite{hetero_PNAS} we have pointed out that population heterogeneity is a dynamic property that roams across multiple timescales. A strong short-term overdispersion of the individual infectivity  manifests itself in the statistics of   super-spreading events. At the other end of the spectrum is a much weaker persistent heterogeneity operating on very long timescales. In particular, it is this long-term heterogeneity that leads to a reduction of the HIT compared to that predicted by classical homogeneous models \cite{Herd_Gomes,hetero_PNAS,neipel2020power, weitz2020heterogeneity,Herd_Science_2020}. 
However, the epidemic dynamics is also sensitive to transient timescales over which the bursty short-term social activity of each individual crosses over to its long-term average.   By including the effects of dynamic heterogeneity, 
we demonstrate  that a suppression of the early waves of the COVID-19 epidemic, even without active mitigation, does not signal achievement of long-term herd immunity. Instead, it is  associated with Transient Collective Immunity, a fragile state which  degrades over time as individuals change their social activity patterns \cite{hetero_PNAS}. 
As we demonstrate below, the first wave is generally followed either by secondary waves or by long plateaus characterized by a nearly constant incidence rate. In the context of COVID-19, both long plateaus and multi-wave epidemic dynamics have been commonly observed \cite{DIVOC}. According to our analysis, the number of daily infections during the plateau regime,  as well as the individual wave trajectories, are robust properties of the epidemic and depend on the current level of mitigation, degree of heterogeneity and temporal correlations of individual social activity.

Our work implies that, once the  plateau-like dynamics is established,  the epidemic gradually evolves towards the long-term HIT determined by persistent population heterogeneity. However, reaching that state may stretch over a surprisingly long time, from months to years. On these long timescales, both waning of individual biological immunity and mutations of the pathogen become  valid concerns, and would ultimately result in a permanent endemic state of the infection. Such endemic behavior is a well-known property of most classical epidemiological models \cite{keeling2011modeling}. However,  the emergence of the endemic state for a newly introduced pathogen is far from being completely understood \cite{emergence_nature2007,emergence_2013,Pastor_endemic}. Indeed, most epidemiological models would typically predict complete extinction of a pathogen following the first wave of the epidemic, well before the pool of susceptible population would be replenished. A commonly accepted, though mostly qualitative, explanation for the onset of endemic behavior of such diseases as measles, seasonal cold, etc., involves geographic heterogeneity: the pathogen may survive in other geographic locations until returning to a hard hit area with a depleted susceptible pool \cite{emergence_nature2007,emergence_2013}.  In contrast, our theory provides a simple and general mechanism  that prevents an overshoot of the epidemic dynamics and thus naturally and generically leads to the endemic fixed point. 

The importance of temporal effects has been long recognized in the context of network-based epidemiological models \cite{Barrat_proximity_2017,dynamic_volz2007susceptible,bansal2010dynamic,Read_dynamical}.  On one hand, available high-resolution data on real-world
temporal contact networks allows direct modeling of epidemic spread on those networks. On the other hand, building upon successes of epidemic models  on static unweighted networks \cite{Lloyd_May_Science_2001,may2001infection,Moreno2002,pastor2015epidemic}, a variety of their  temporal generalizations have been proposed. Those typically involve particular  rules for  discrete or continuous network rewiring \cite{dynamic_volz2007susceptible,bansal2010dynamic,Read_dynamical}
such as e.g. in  activity-based network models \cite{Perra2012activity,Barabasi2007PRL_activity, behavior_activity}. While important theoretical results have been obtained for some of these problems, especially regarding the  epidemic threshold, many open questions and challenges remain in the field. In this paper,  we  start with a more traditional heterogeneous well-mixed model, which is essentially  equivalent to the  mean-field description of an epidemic on a network \cite{Moreno2002,Pastor_Scale_free,bansal2007individual}, and include effects of time-variable social activity that modulates levels of individual susceptibilities and infectivities.

\begin{figure}[ht]
\centering
\includegraphics[width=1\columnwidth]{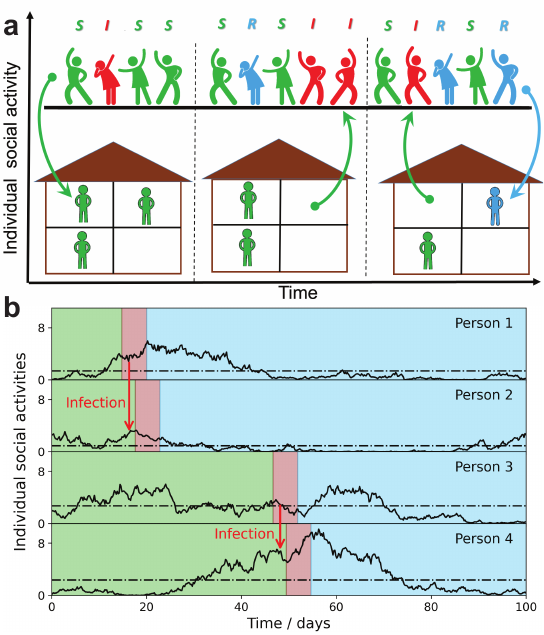}
\caption{Schematic illustration of our model in which each individual is characterized by a time-dependent social activity. a)
People with low social activity (depicted as socially isolated figures at home) occasionally increase their level of activity (depicted as a party). The average activity in the population remains the same, but individuals constantly change their activity levels from low to high (arrows pointing up) and back (arrows pointing down). Individuals are colored according to their state in the SIR epidemiological model: susceptible - green, infected - red, and removed - blue. The epidemic is fuelled by constant replenishment of susceptible population with high activity due to transitions from the low activity state. b) examples of  individual time-dependent activity $a_i(t)$ (solid lines), with different persistent levels $\alpha_i$ (dot-dashed lines). S,I,R states of an individual  have the same color code as in (a). Note that pathogen transmission  occurs predominantly between individuals with high current activity levels.
}
\label{fig:fig1}
\end{figure}
The basic idea behind our model is represented in  Fig. \ref{fig:fig1}. Each individual $i$ is characterized by time-dependent social activity $a_i(t)$ proportional to his/her current frequency and intensity of close social contacts. This quantity determines both individual susceptibility to infection as well as ability to infect others. The time evolution of contact frequency, and hence $a_i(t)$,  is in principle measurable by means of proximity devices, such as RFID, Bluetooth, Wi-Fi, etc. \cite{proxi_PNAS2010,Barrat_proximity_2017,Crowd,pastor2015epidemic}.   
%
%
%
In our model we combine a simple mathematical description of social dynamics with the standard Susceptible-Infected-Removed (SIR) epidemiological model. Qualitatively it leads to long-term epidemic dynamics fuelled by replenishment of susceptible population due to changes in the level of individual social activity from low to high.
Fig. \ref{fig:fig1}(a) illustrates this process by showing people with low social activity (depicted as socially isolated at home) occasionally increasing their level of activity (depicted as a party). Fig. \ref{fig:fig1}(b) represents the same dynamics in terms of individual functions $a_i(t)$. Note that each person is characterized by his/her own long-term average activity level $\alpha_i$ (dot-dashed lines), but  the transmission occurs predominantly between individuals with high levels of {\it current} social activity.  This is because  $a_i(t)$ determines both current susceptibility and individual infectivity of a person. However, the secondary transmission is delayed with respect to the moment  of infection, by a time of the order of a single generation interval $\tau_g$ (around 5 days for COVID-19). Studies of real-world contact and interpersonal communication networks have shown that individual social activity is bursty and varies across multiple timescales---from seconds to years \cite{rybski2009scaling,barabasi2005origin,saramaki2015seconds,Sneppen2020}.   

For any individual $i$ the value of $a_i(t)$ has a tendency to gradually drift towards its persistent average level $\alpha_i$, which itself varies within the population. In  our model, we assign a single timescale $\tau_s $ to this mean reversion process. This is of course a simplification of the multi-scale relaxation observed in real social dynamics. While  $\tau_s$ can be treated as a fitting parameter of our model, here we simply set it to be $\tau_s=30$ days, several fold longer than the mean generation  interval of COVID-19, $\tau_g=5$ days. 
Note that from the point of view of the epidemic dynamics, variations in activity on timescales shorter than the mean generation interval may be safely ignored. For example, attending a single party would increase an individual's risk of infection but would not change his/her likelihood of transmission to others 5 days later. 

The individual social activity $a_i(t)$ is governed by the following stochastic equation:
\begin{equation}
 \label{Langevin}
     \dot{a_i}(t)=\frac{\alpha_i-a_i(t)}{\tau_s}+\eta_i(t)
\end{equation}
Here $\eta(t)$ is a short-time noise that gives rise to time-dependent variations in $a_i(t)$.  We set $\langle\eta_i(t)\eta_i(t')\rangle=\frac{2a_i(t)}{\tau_s k_0}\delta(t-t')$, which results in a diffusion in the space of individual social activity with diffusion coefficient proportional to $a_i$. This stochastic process is well known in mathematical finance as the Cox–Ingersoll–Ross (CIR) model \cite{CIR} and has been studied in probability theory since 1950s \cite{Feller}. The major properties of this model are (i) reversion to the mean and (ii) non-negativity of $a_i$ at all times, both of which are natural for social activity. Furthermore, the steady state solution of this model is characterized by the Gamma-distributed $a_i$. This is consistent with the empirical statistics of short-term overdispersion of disease transmission manifesting itself in superspreading events \cite{LloydSmith2005,Super_Kucharski,Vespigniani2020}.
More specifically, for a given level of persistent activity  $\alpha$,  this model generates the steady-state distribution  of ``instantaneous'' values of  social activity $a$ following gamma distribution with mean $\alpha$ and variance $\alpha/k_0$: $f_\alpha(a) \sim a^{k_0\alpha-1}e^{-k_0a}$. 

The statistics of super-spreader events is usually represented as a negative binomial distribution, derived  from a gamma-distributed individual reproduction number \cite{LloydSmith2005}. The observed overdispersion parameter $k\approx 0.1-0.3$ \cite{Super_Kucharski,Vespigniani2020} can be used for partial calibration of our model. This short-term overdispersion has both  stochastic and persistent contributions. In our model, the former is characterized by dispersion $k_0$. In addition, we assume  persistent levels of social activity $\alpha_i$ to also  follow a Gamma distribution with another dispersion parameter,  $\kappa$.  In several recent studies  of epidemic dynamics in populations with persistent heterogeneity \cite{hetero_PNAS, Aguas_Gomes_Hetero_2020,neipel2020power} it has been demonstrated that  $\kappa$ determines the herd immunity threshold. Multiple studies of real-world contact networks  (summarized, e.g. in \cite{bansal2007individual})  report an approximately exponential distribution of $\alpha$, which  corresponds to $\kappa \simeq 1$. Throughout this paper, we assume a more conservative value, $\kappa = 2$, i.e. coefficient of variation $1/\kappa=0.5$, half way between the fully homogeneous case and that with exponentially distributed $\alpha$.  For consistency with the reported value of the short-term  overdispersion parameter \cite{Vespigniani2020}, $1/k\approx 1/\kappa+1/k_0\approx 3$, we set $k_0=0.4$.  
%
%
%

According to Eq. (\ref{Langevin}),  individuals, each characterized by his/her own persistent level of social activity $\alpha$, effectively diffuse in the space of their current social activity $a$. This leads to major modifications of the epidemic dynamics. For instance, the equation for the  susceptible fraction in classical epidemic models \cite{keeling2011modeling} acquires the following form:
\begin{equation}
  \label{eq:S} \dot {S}_\alpha(a,t)=\left[-a J(t)  + \frac{a}{k_0\tau_s}  \frac{\partial^2}{\partial a^2} + \frac{\alpha-a}{\tau_s}\frac{\partial }{\partial a}\right] S_\alpha(a,t)
\end{equation}
Here $S_\alpha(a,t)$ is the fraction of susceptible  individuals within a subpopulation with a given value of persistent social activity $\alpha$ and with current social activity $a$,  at the moment of infection, and $J(t)$ is the current strength of infection. Its time evolution can be described by any traditional epidemiological model, such as e.g. age-of-infection, SIR/SEIR, etc \cite{keeling2011modeling}.  

\begin{figure}[ht]
\centering
\includegraphics[width=1\columnwidth]{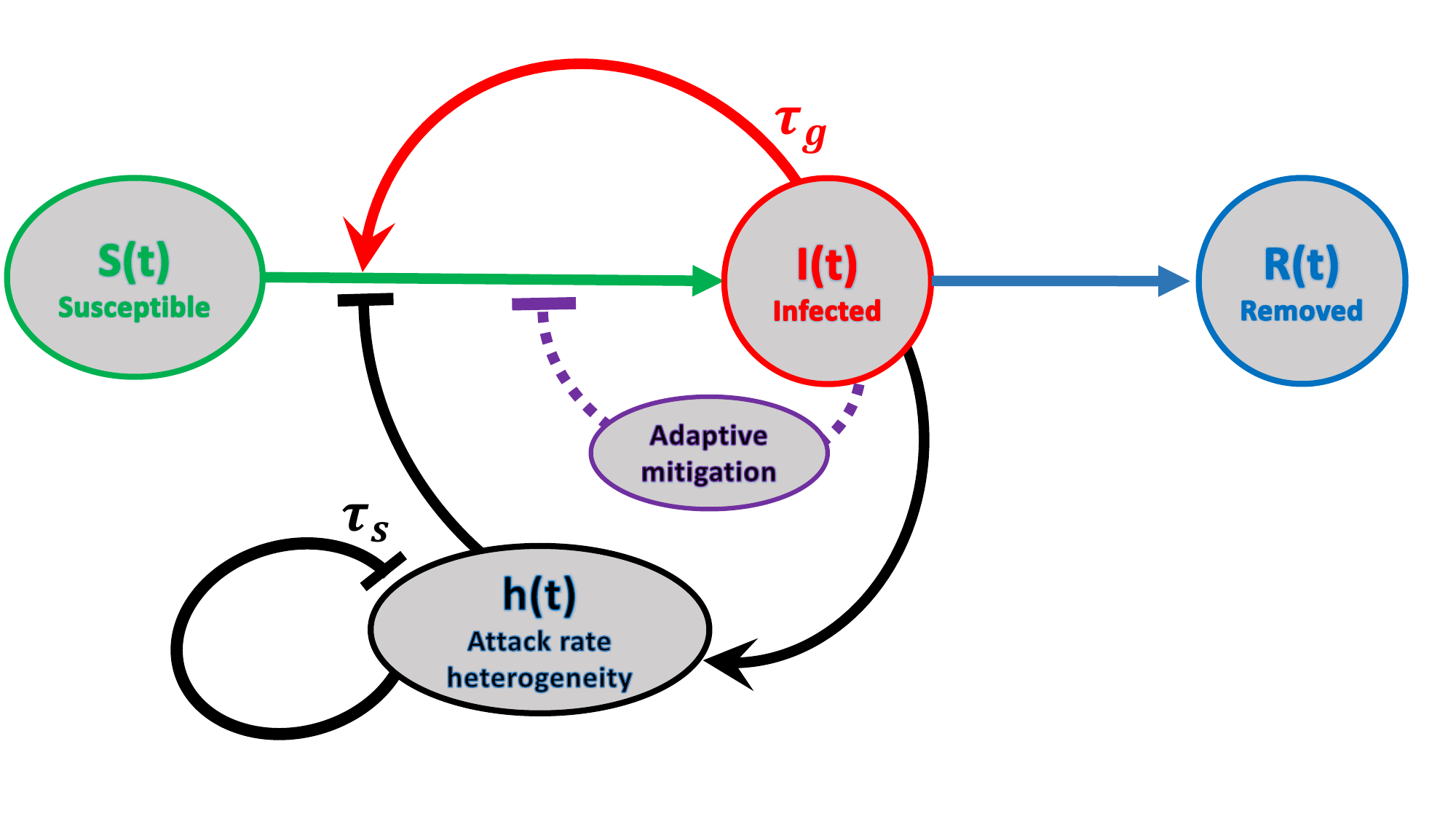}
\caption{Schematic representation of the mechanisms that lead to  self-limited epidemic dynamics. In traditional epidemic models, the major factor  is the depletion  of susceptible population. Government-imposed mitigation and/or behavioral  adaptation to the perceived risk create another feedback loop (purple). Yet another mechanism, is due to the dynamic heterogeneity of the attack rate parameterized by $h(t)$ (black). On the one hand, the attack rate heterogeneity is being generated by the current infection. On the other hand, it suppresses itself on the timescale of $\tau_s$ due the reversion of individual social activity towards the mean (black feedback loop). 
}
\label{fig:diagram}
\end{figure}

Eq. (\ref{eq:S}), is dramatically simplified by writing $S_\alpha(a,t)\equiv e^{-Z(t)\alpha- k_0h(t)a}$, implicitly defining $Z(t)$ as a measure of persistent heterogeneity of the attack rate and the transient heterogeneity parameter $h(t)$. As $Z$ increases, so does the difference in depletion of susceptibles among subpopulations with different $\alpha$'s, i.e. various levels of persistent social activity. On the other hand, $h(t)$ parametrizes the transient heterogeneity within each of these subpopulations. Both $Z(t)$ and $h(t)$ indicate the current level of heterogeneity of the attack rate i.e. susceptible population structure. In the long run, transient heterogeneity disappears due to the diffusion in the $a$-space, thus $h(t)$ asymptotically approaches $0$ as $t \to \infty$. We combine this ansatz with a general methodology \cite{hetero_PNAS} that provides a quasi-homogeneous description for  a  wide variety of heterogeneous epidemiological models.  For a specific case of SIR  dynamics, we assign  each  person a state variable $I_i$ set to  $1$ when the individual is infectious   and $0$ otherwise. Now, the activity-weighted fraction of the infected population is  defined as    $I(t)=\langle I_ia_i(t)\rangle/\langle a_i^2\rangle$, and the current infection strength is proportional to it:    $J(t)=R_0M(t)I(t)/\tau_g$. Here $M(t)$ is a time-dependent mitigation factor,  which combines the effects of  government interventions, societal response to the epidemic, as well other sources of time modulation, such as e.g. seasonal forcing. 

Using this ansatz, the epidemic in a population with both persistent and dynamic heterogeneity of individual social activity can be compactly described as a dynamical system with only three  variables:  the susceptible population fraction $S(t)$, the infected population fraction $I(t)$ (activity-weighted) which is proportional to strength of infection $J(t)$,  and the transient heterogeneity parameter $h(t)$.    In the  $(S,I,h)$-space, the dynamics  is given by the following set of differential equations:
\begin{equation}
     \label{eq:JSH1_main}
     \frac{dI}{dt}=  \frac{J S ^\lambda}{\left(1+h\right)^2}- \frac{I}{\tau_g}  
     \end{equation}
     \begin{equation}
    \label{eq:JSH2_main} \frac{dS}{dt}= -\frac{JS^{1+1/\kappa}}{(1+h)} \\ 
      \end{equation}
     \begin{equation}
    \label{eq:JSH3_main} \frac{dh}{dt}=\frac{J}{k_0}-\frac{h(1+h)}{\tau_s}
   \end{equation}
As in the case of persistent  heterogeneity  without temporal variations \cite{hetero_PNAS}, the long-term herd immunity threshold,  $1-(R_0M)^{-1/\lambda}$,  is determined by the immunity factor $\lambda$. The latter depends both on the  short-term and persistent dispersion parameters:
\begin{equation}
    \lambda=\frac{\left(1+\kappa^{-1}\right)\left(1+k_g^{-1}+2\kappa^{-1}\right)}{1+k_g^{-1}+\kappa^{-1}}
\end{equation}
Here $k_g=k_0(1+\tau_g/\tau_s)$ is the dispersion parameter for  activity $a(t)$ averaged over a timescale of a single generation interval. 
For parameters $k_0=0.4$ and $\kappa=2$, $\tau_s=30$ days used throughout  our study, $k_g=0.47$, $\lambda=1.7$, consistent with our earlier estimate of  $\lambda_{\infty} \approx 2$ \cite{hetero_PNAS}. 
%
%
%

\begin{figure}[ht]
\centering
\includegraphics[width=1\columnwidth]{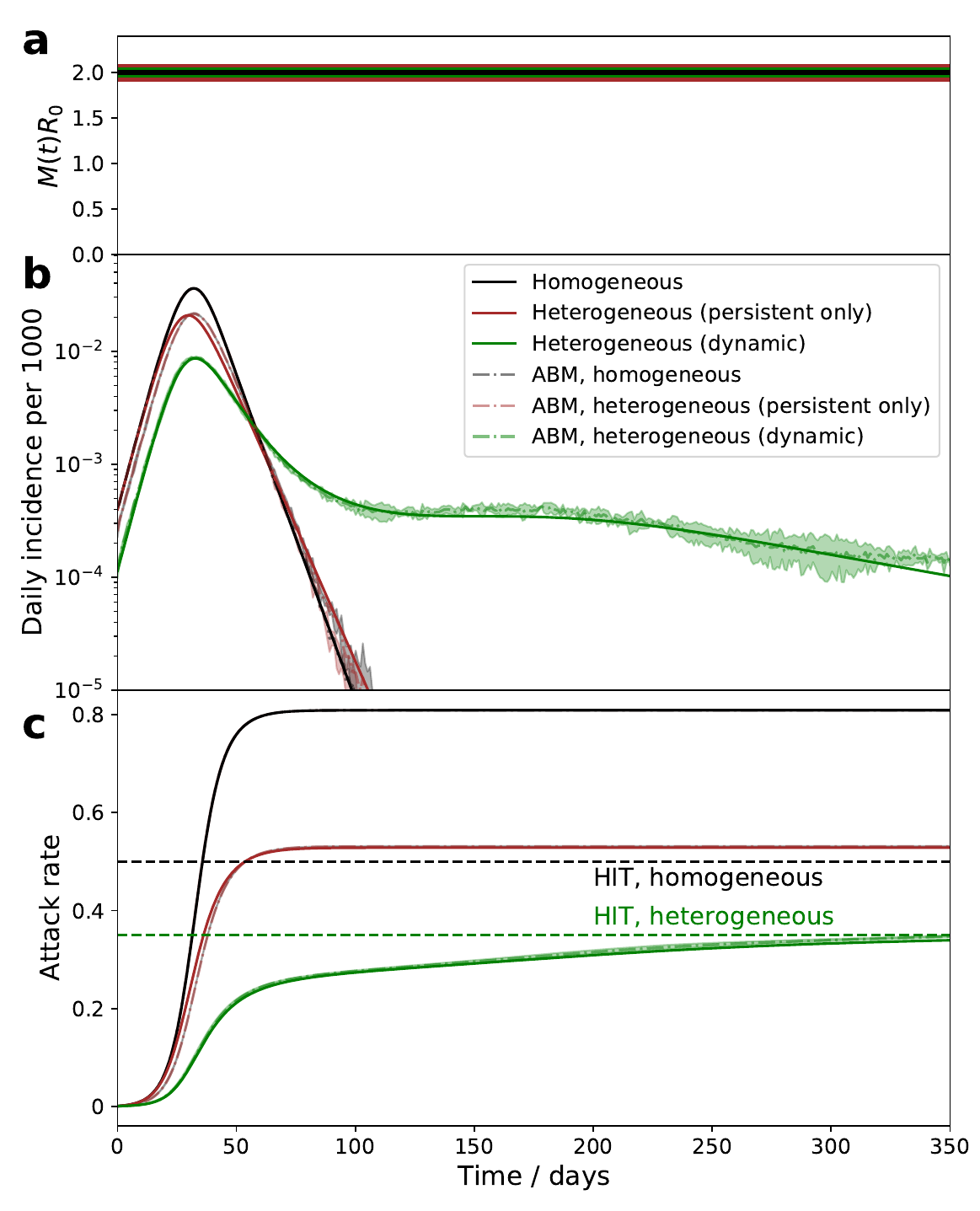}
\caption{  (a)-(c)   Comparison of the epidemic dynamics with homogeneous population (black curves), persistent population heterogeneity (brown curves), and 
with dynamic heterogeneity (green curves): mitigation profile (a), daily incidence (b), and cumulative attack rate (c).  
While parameters in cases (b) and (c) correspond to the same herd immunity threshold (HIT), the behavior is drastically different. In the persistent model, the epidemic quickly overshoots above HIT level. In the case of dynamic heterogeneity, the initial wave is followed by a plateau-like behavior with  slow relaxation towards the HIT.  Note an excellent agreement between the quasi-homogeneous theory described by Eqs. (\ref{eq:JSH1_main}- \ref{eq:JSH3_main}) (solid lines) and the Agent-Based Model with 1 million agents whose stochastic activity is given by Eq. (\ref{Langevin}) (shaded area $=$ the range of 3 independent simulations).}  
\label{fig:epidemic_curves}
\end{figure}

\begin{figure}[ht]
\centering
\includegraphics[width=1\columnwidth]{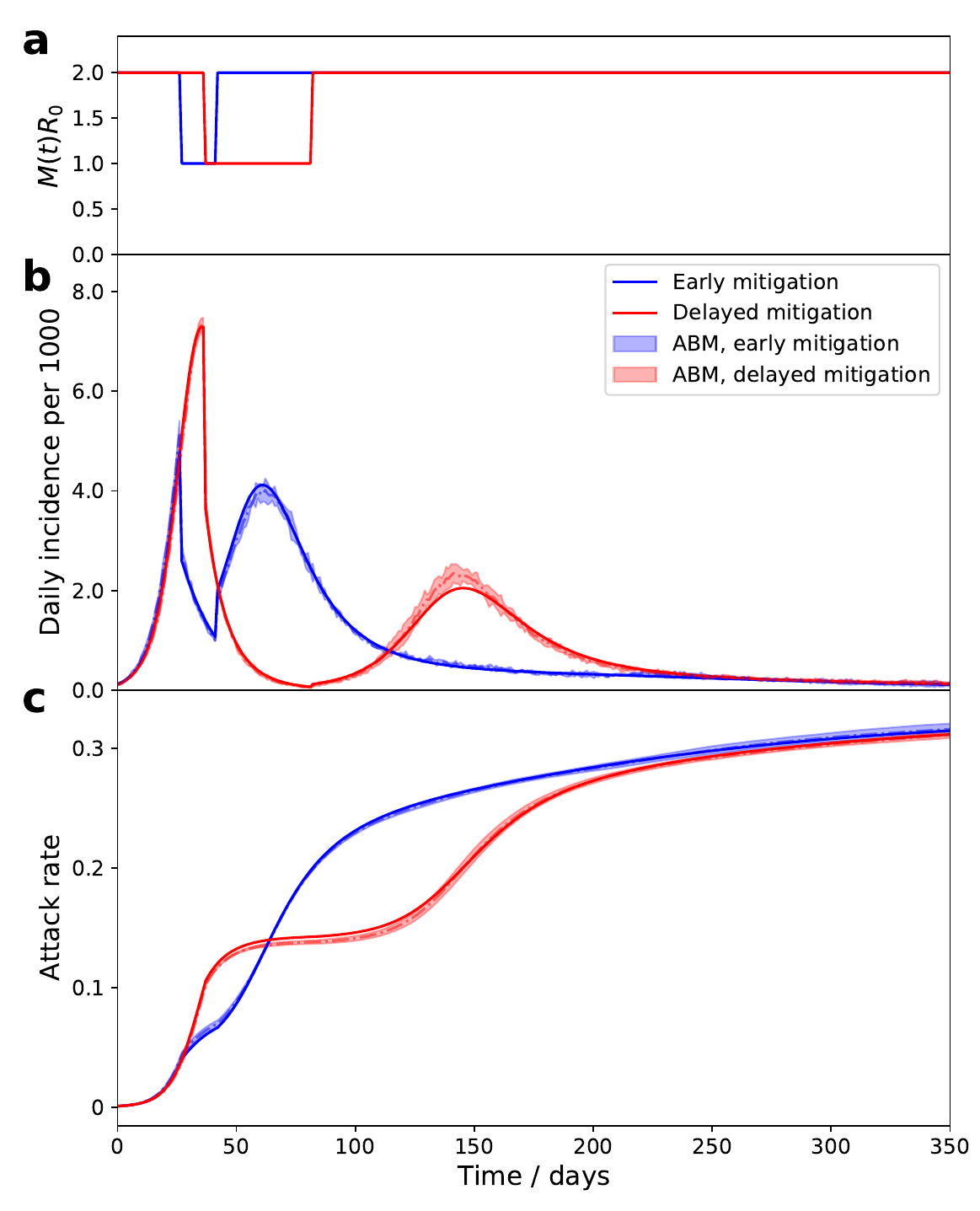}
\caption{The time course of an epidemic  with  enhanced mitigation during the first wave.  (a) shows the  $M(t)R_0$ progression for two different strategies.  In both cases, the enhanced mitigation leads to a 50\% reduction of $M(t)R_0$ from $2$ to $1$. In the first scenario 
(early mitigation, blue curves), the reduction lasted for only $15$ days starting from day $27$. In the second scenario 
(delayed mitigation, red curves), the mitigation was applied on day $37$ and lasted for $45$ days. (b)-(c) show daily incidence and cumulative attack rates for both strategies.  As predicted, differences in the initial mitigation had no significant effect on the epidemic in the long run: the two trajectories eventually converge towards the universal attractor.  However, the early mitigation allows to suppress the peak of the infection, potentially reducing the stress on healthcare system. A delayed mitigation gives rose to a sizable second wave.}
\label{fig:mitigated_epidemic_curves}
\end{figure}

In Figure \ref{fig:diagram} we schematically represent three feedback mechanisms that lead to self-limited epidemic dynamics. The most conventional of them relies on depletion of susceptible population (red). Another mechanism is due to government mitigation as well as personal behavioral response to perceived epidemic risk (purple). Finally, according to our theory there is yet another generic mechanism related to accumulated  heterogeneity of the attack rate, quantified by parameter   $h(t)$.  Due to the long-term relaxation of $h(t)$ this feedback loop limits the scale of a single epidemic wave, but does not provide long-term protection against new ones. 

As demonstrated below, the theory described by Eqs. (\ref{eq:JSH1_main})-(\ref{eq:JSH3_main}) is in excellent agreement with simulations of the Agent-Based Model (ABM) in which social activities of 1 million agents undergo stochastic evolution described by Eq. (\ref{Langevin}) (compare solid lines with shaded areas in Figs. \ref{fig:epidemic_curves}, \ref{fig:mitigated_epidemic_curves}).

Figure \ref{fig:epidemic_curves} illustrates a dramatic effect time-dependent heterogeneity has on the epidemic dynamics. It compares three cases: the classical homogeneous SIR model (black), the same model with persistent heterogeneity (brown), and the dynamic heterogeneity case  considered in this study (green). The latter two models share the same HIT (green dashed line) which is reduced compared to the homogeneous case (black dashed line). In the absence of dynamic heterogeneity (black and brown) the initial exponential growth halts once the respective HIT is reached, but the overall attack rate 
``overshoots'' beyond that point, eventually reaching a significantly larger level, known as the  final size of the epidemic (FSE).  Importantly, in both these cases the epidemic has only a single wave of duration set by the mean generation interval $\tau_g$ multiplied by a certain   $R_0$-dependent factor. 
In the case of dynamic heterogeneity (green), described by Eqs. (\ref{eq:JSH1_main})-(\ref{eq:JSH3_main}), the epidemic is transiently suppressed at the level which is below even the heterogeneous HIT. As we argued in Ref. \cite{hetero_PNAS} this temporary suppression is due to the population reaching the Transient Collective Immunity (TCI). That state originates due to the short-term population heterogeneity being enhanced compared to its persistent level.  Stochastic contributions to social activity responsible for this enhancement eventually average out, leading to a slow degradation of the TCI state.   
Fig. \ref{fig:epidemic_curves}b illustrates that as the TCI state degrades, the daily incidence rate develops an extended plateau on the green curve.  The cumulative attack rate shown in Fig. \ref{fig:epidemic_curves}c  relaxes towards the HIT. This relaxation is characterized by an emergent long time constant $\tilde{\tau}\simeq \tau_s/k_0 > \tau_s$.

According to Eqs. (\ref{eq:JSH1_main}-\ref{eq:JSH3_main}) for a fixed mitigation level $M(t)$, any epidemic trajectory would eventually converge to the same curve, i.e. the universal attractor. The existence of the universal attractor  
is apparent in Fig \ref{fig:mitigated_epidemic_curves}, where  we compare two scenarios with different mitigation strategies applied at early stages of the epidemic. In both cases, an enhanced mitigation was imposed leading to a reduction of $M(t)R_0$ by $50\%$ from $2$ to $1$. In the first scenario 
(blue curves), the enhanced mitigation was imposed on day $27$ and lasted for $15$ days. In the second scenario 
(red curves), the mitigation was applied on day $37$ and lasted for $45$ days. As predicted, this difference in mitigation has not had any significant effect on the epidemic in the long run: these two trajectories eventually converged towards the universal attractor. However, short- and medium- term effects were substantial. The early mitigation scenario (blue curve) resulted in a substantial suppression of the maximum incidence during the first wave.
Immediately following the release of the mitigation the second wave started and reached approximately the same peak value as the first one. If the objective of the intervention  is to avoid the overflow of the healthcare system, this strategy would indeed help to achieve it. 
In contrast, the delayed mitigation scenario (red curve) turned out to be largely counterproductive. It did not suppress the peak of the first wave, but brought the infection to a very low level after it.
Eventually, that  suppression backfired as the TCI state  deteriorated and  the epidemic resumed as a second wave, which is not as strong as the first one.

Since the late-stage evolution in our model is characterized by a long relaxation time $\tilde{\tau}$, the possibility of waning of individual biological immunity or escape mutations of the pathogen accumulated over certain (presumably, also long)  time  $\tau_b$,   becomes a relevant effect.  It can be incorporated as an additional relaxation term 
$(1-S)/\tau_b$ in  Eq. (\ref{eq:JSH2_main}). The analysis of our equations, modified in this way,  shows  that the universal attractor leads to a fixed point  corresponding to the  endemic state. That point is located somewhat below the HIT and characterized by the finite residual incidence rate $(1-S_\infty)/\tau_b$ and, respectively, by finite values of $I$ and $h$. Here $S_\infty$ is a susceptible population fraction in the endemic state, which is close to, but somewhat higher than that at the onset of the herd immunity. A similar endemic steady state exists in most classical epidemic models (See \cite{keeling2011modeling} and references therein). However, in those cases, the epidemic dynamics would not normally lead to that point  due to the overshoot phenomenon. Instead, those models typically predict a complete extinction of the disease when the prevalence drops below one infected individual. This may happen before  herd immunity is lost due to waning  biological immunity and/or replenishment of the susceptible population (e.g. due to births of immunologically naive individuals). That is not the case when the time-dependent heterogeneity is included. 

\begin{figure}[ht]
\includegraphics[width=1\columnwidth]{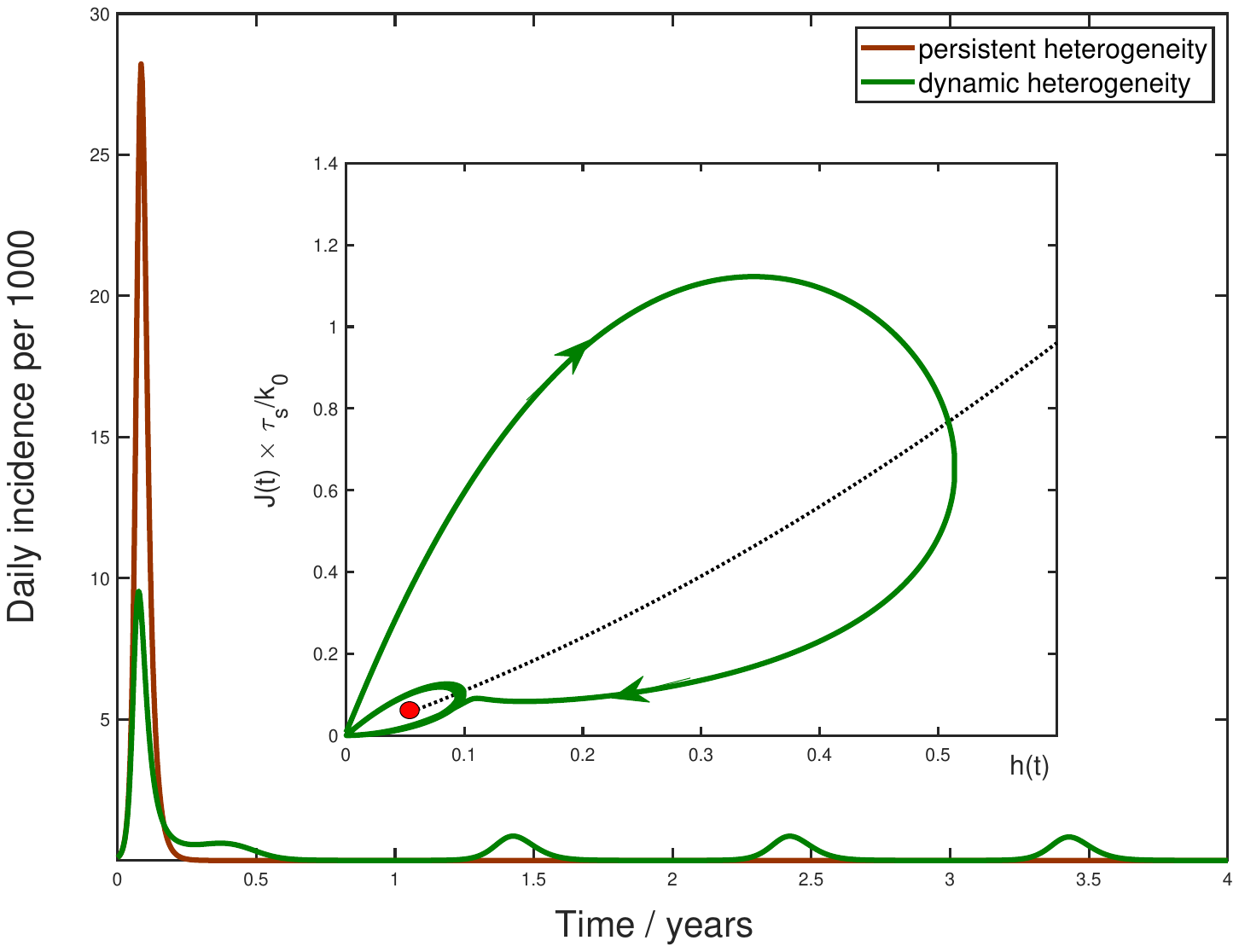}
\caption{Multi-year dynamics of a new pathogen. Effects of waning biological immunity with characteristic time $\tau_b=5$ yrs, and seasonal forcing are included (see SI for details). In the case of  persistent heterogeneity without temporal variations of social activity (brown solid line) the infection gets extinct following the initial wave of the epidemic. In contrast, dynamic heterogeneity leads to an endemic state with strong seasonal oscillations  (green line). Insert: The epidemic dynamics in the $(J,h)$ phase space. The black dotted line corresponds to the universal attractor trajectory manifested e.g. as a plateau in green line in Fig. \ref{fig:epidemic_curves}b. The attractor leads to the endemic state (red point). }
\label{fig:phase_diagram}
 \end{figure}

Note that for most pathogens the endemic point is not fixed but instead is subjected to periodic seasonal forcing in $M(t)$. This leads to  annual peaks and troughs in the incidence rate. Our model is able to describe this seasonal dynamics as well as transition towards it for a new pathogen (see Fig. \ref{fig:phase_diagram}). It captures the important qualitative features of seasonal waves of real pathogens, e.g. three endemic coronavirus families studied in Ref. \cite{Neher2020}. They are (i) sharp peaks followed by a prolonged relaxation towards the annual minimum; (ii) a possibility of multi-annual cycles due to parametric resonance. 

To understand the nature of the overall epidemic dynamics, we focus on  the behavior of variables $J(t)$ and $h(t)$. Their evolution is described by  Eqs.(\ref{eq:JSH1_main}) and (\ref{eq:JSH3_main}),  with  $R^*=R_0M(t) S(t)^\lambda$  playing the role of a driving force.  As a result of depletion of susceptible population, the driving force is gradually reduced, and the dynamic converges  towards a  slow evolution along the universal attractor shown as a black dotted trajectory  in $(h,J)$ coordinates at the  insert to Fig. \ref{fig:phase_diagram}. For initial conditions away from that trajectory (say, $J\approx 0$, $h=0$), the  linear stability analysis indicates  that the epidemic dynamics has a damped oscillatory behavior manifesting itself as a spiral-like relaxation towards the universal attractor. A combination of this spiral dynamics with a slow drift towards the endemic state gives rise to the overall trajectory shown as the solid green line at the  insert to  Fig. \ref{fig:phase_diagram}. The periodic seasonal forcing generates a limit cycle about the endemic point (small green ellipse around the red point). 
 
More generally, any abrupt increase of the effective reproduction number e.g. due to a relaxed mitigation, seasonal changes, etc. would shift the endemic fixed point up along the universal attractor. 
According to Eqs. (\ref{eq:JSH1_main}-\ref{eq:JSH3_main}) this will once again trigger a spiral-like relaxation. It will manifest itself as a new wave of the epidemic, such as the secondary waves in Fig. \ref{fig:mitigated_epidemic_curves}b).

\begin{figure}[ht]
\includegraphics[width=1\columnwidth]{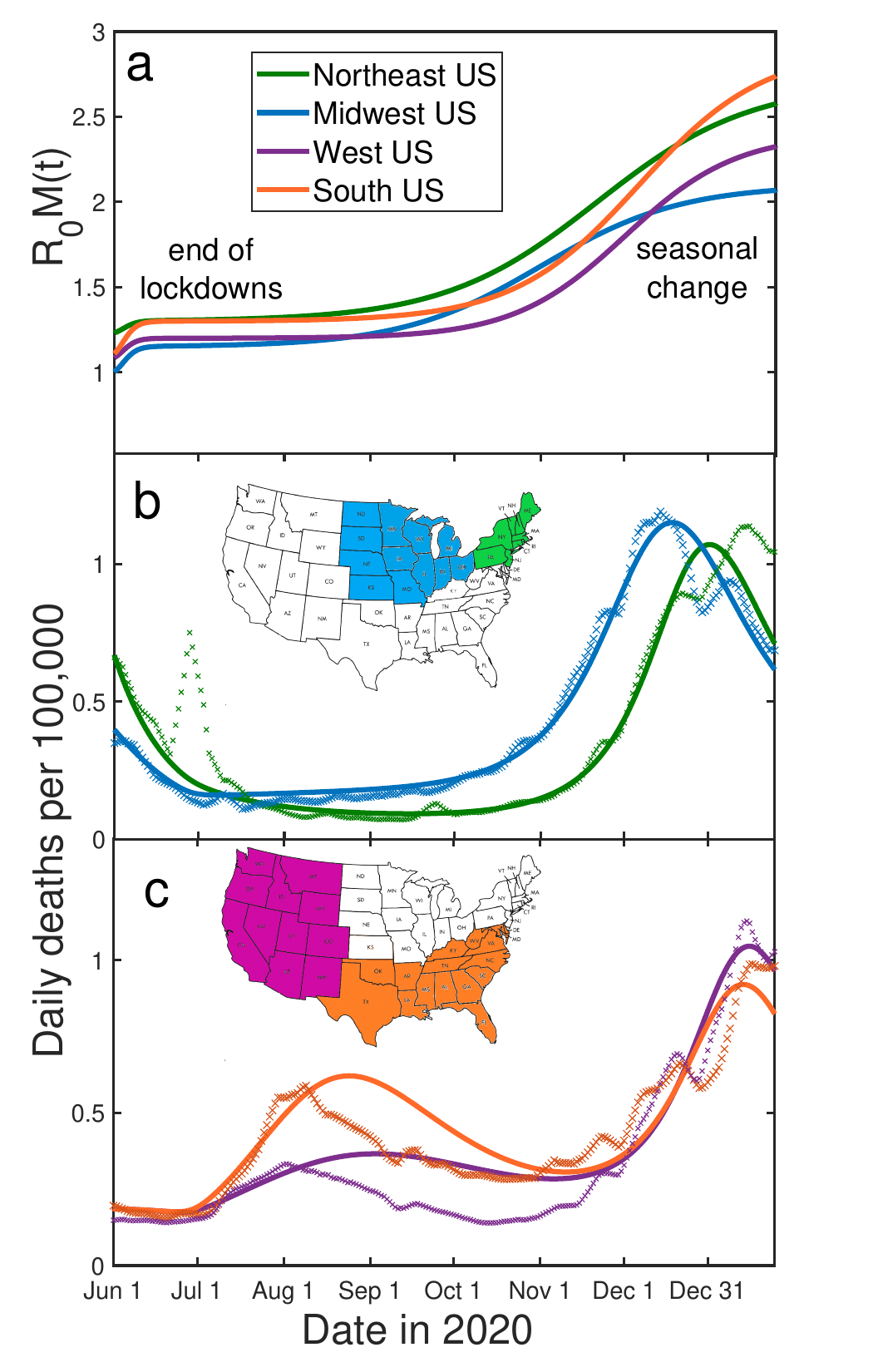}
\caption{Fitting of the empirical data on COVID-19 epidemic in Northeast (green),  Midwest (blue), West (purple) and South (orange) of the USA.  
(a) The best-fit  profiles  $R_0M(t)$, under the assumption that they are shaped by the end of lockdowns in the early summer of 2020, followed by gradual  seasonal changes in the fall.  Time dependence of daily deaths per capita for the Northeast and  Midwest as well as  for  South and  and  West US are shown in panels (b) and (c), respectively.   Data points represent  reported daily deaths per 100,000 of population for each of the regions \cite{DIVOC}. Solid lines are the theoretical fits with our model.  The following parameters of $M(t)R_0$ curves  were subject to variation: the  mitigation level $M(t)$ before and after the termination of the lockdown, the amplitude of the seasonal forcing, and the summer-winter crossover time.   Other  parameters have been fixed:   $k_0=0.4$, $\kappa=2$,   $\tau_s=30$ days, $IFR=0.4\%$, and $\tau_b=5$ years.  }
\label{fig:southwest}
 \end{figure}

A systematic calibration of our model and application to real-world epidemiological data is beyond the scope of this study. However, below we present a proof-of-principle demonstration that the progression in the COVID-19 epidemic in the summer and fall of 2020 in four regions of the continental US: South, Northeast, Midwest and West can be well described by our theory. The time dependence of daily deaths per capita (a reliable, albeit delayed measure proportional to the true attack rate) is  shown in Fig. \ref{fig:southwest}bc for each of the regions and fitted by our model with $k_0=0.4$, $\tau_s=30$ days, $\kappa=2$, together with IFR assumed to be $0.4\%$. This IFR value was estimated by comparing reported COVID-related  deaths in USA  \cite{DIVOC} to two independent seroprevalence surveys \cite{sero1, sero2}. 
%
%
%
%
Unlike during the first wave, state-mandated mitigation measures have been relaxed in the early summer and kept largely constant during this time. Thus it is reasonable to assume that $M(t)$ follows regular seasonal dynamics during that time period. This is reflected in a simple mitigation profile  $R_0M(t)$ shown in Fig. \ref{fig:southwest}a featuring a sharp relaxation of  mitigation in early summer 2020 and gradual seasonal increase of the reproduction number during the fall. Release of mitigation triggered an immediate second wave in the South and West (Fig. \ref{fig:southwest}c) that had relatively low exposure during the first wave of the epidemic. In our model the wave is transiently suppressed when these regions reach the TCI state. Differences between the data and our model observed in the early fall may be tentatively attributed to government-imposed mitigation measures and/or to knowledge-based adaptation of the human behavior \cite{PNAS2009_behavior,behavior_activity,Weitz2020_awareness}. The late fall wave in all regions was triggered by seasonal changes in transmission. According to our model this wave was stabilized in mid-winter due to the population once again reaching the TCI state. Note a good agreement between peak levels of this wave with our predictions.

Midwest and Northeast (Fig. \ref{fig:southwest}b) exhibited similar behavior except having a plateau instead of the summer wave due to their higher levels of exposure during the first wave of the epidemic in spring 2020.
Importantly, the transmission monotonically increases throughout the entire time window, yet our model captures the observed secondary waves and plateaus. That behavior would be impossible to explain using traditional epidemiological models well below the herd immunity. 

In conclusion, we have proposed a new theory integrating the stochastic dynamics of individual social activity into traditional epidemiological models. Our model describes the so-called ``zero intelligence'' limit in which there is no feedback from the epidemic dynamics to social activity e.g. mediated by the news. Hence our approach is complementary to knowledge-based models of Refs. \cite{PNAS2009_behavior,behavior_activity,Weitz2020_awareness}. The stochastic social activity in our approach is described by the CIR model \cite{CIR} which captures the  following important properties: (i) the activity cannot be negative; (ii) for any given individual it reverses towards its long-term average value; (iii) it exhibits gamma-distributed short-term overdispersion (aka superspreading) \cite{LloydSmith2005,Super_Kucharski,Vespigniani2020}. We mapped the overall epidemic dynamics featuring heterogeneous time-varying social activity onto a system of three differential equations, two of which generalize the traditional SIR model. The third equation describes the dynamics of the heterogeneity parameter $h(t)$, driven up by the current strength of infection $J(t)$ and relaxing back to zero due to variable social activity. 

The emergent property of our theory is the new long timescale $\tau_s/k_0$ governing the relaxation towards either the herd immunity or the endemic state of the pathogen. This behavior is in striking contrast to traditional epidemiological models generally characterized by a large overshoot beyond the herd immunity threshold leading to a likely extinction of new pathogens. Our theory is in a good agreement with the empirical observation of long plateaus observed for many real-life epidemics including COVID-19 \cite{DIVOC}.   Dynamic heterogeneity also leads to a transient suppression of individual waves of the epidemic without reaching the long-term herd immunity \cite{hetero_PNAS}. 
Finally, we demonstrated that our theory is in quantitative agreement with the data describing secondary waves of COVID-19 epidemic in different regions of the USA.

\begin{acknowledgments}
This work was supported by the University of
Illinois System Office, the Office of the Vice-Chancellor for Research
and Innovation, the Grainger College of Engineering, and the Department
of Physics at the University of Illinois at Urbana-Champaign.
This research was partially done at, and used resources of the Center for Functional Nanomaterials, which is a U.S. DOE Office of Science Facility, at Brookhaven National Laboratory under Contract No.~DE-SC0012704.
\end{acknowledgments}

\bibliographystyle{Science}
\bibliography{main}
\clearpage
\onecolumngrid

\newpage
\clearpage
\pagenumbering{arabic}
\setcounter{page}{1}
\renewcommand{\thefigure}{S\arabic{figure}}
\setcounter{figure}{0}
\renewcommand{\theequation}{S\arabic{equation}}
\setcounter{equation}{0}
\section*{Supplementary Materials}

\subsection {Epidemic dynamics with dynamic heterogeneity}
Let $a_i(t)$ be a measure of individual's social activity proportional to frequency and intensity of close contacts with other people around time $t$. We refer to it as  (social) susceptibility to infection, but it also determines one's potential  to infect others. In particular, infectivity  of a person $i$ infected at time $t^*_i$ is given by 
 \begin{equation}
 \label{eq:Ri}
\beta_{i}(t^*_i+\tau) =C_i (\tau) a_i(t^*+\tau)
\end{equation}
Here $C_i(\tau)$ is  person's  contagiousness level, at time $\tau$ after infection.
 
Let $j(t)$  be the fraction of  infected individual, weighted proportionally to their current infectivity level, and   $M(t)$ be mitigation factor that reflects   government and social response to epidemics, seasonal effects etc.  Their product,   $J(t)=M(t)j(t)$ is force of infection,   i.e. a hypothetical incidence rate in fully susceptible homogeneous population with  $\alpha=1$. Within  the heterogeneous (but well-mixed) age-of-infection model, current value of $j(t)$ is given by  
  \begin{equation}
 j(t)= \left \langle   \beta_i(t-t_i^*)\right \rangle_i =\int_{0}^{\infty} \left \langle C_i (\tau) a_i(t) a_i(t-\tau)S_i(t-\tau)  \right \rangle_i J(t-\tau)d\tau 
\end{equation}
Here $S_i(t-\tau)$ is the state of an individual $i$ ($1$ if susceptible, $0$ otherwise). Since  $J(t)$ is by definition proportional to $j(t)$, we obtain   the quasi-homogeneous renewal equation:
\begin{equation}
\label{attack0}
    j(t)=\int_{0}^{\infty} { K(t,\tau)R_e(t-\tau)j(t-\tau) d\tau}
\end{equation}
Here effective reproduction number $R_e$ and the probability density of the generation interval  $K(\tau)$,  are given by
\begin{align}
\label{eq:Re}
R_e(t)&=M(t)\left \langle S_i(t) \int_0^{\infty}{ a_i(t) a_i(t+\tau)C_i (\tau) d\tau}  \right \rangle_i\\
\label{eq:gen_int}
K(t,\tau)&=\frac{\left\langle S_i(t)  a_i(t) a_i(t+\tau)C_i(\tau)\right \rangle_i}{\left\langle S_i(t) \int_0^{\infty}{ a_i(t) a_i(t+\tau)C_i (\tau) d\tau}  \right \rangle_i}
\end{align}

\subsection{Stochastic model for social activity}
It is well known that social interactions are \lq\lq bursty". That is to say,  individual social activity has both (nearly)  permanent and significant time-dependent contributions:
\begin{equation}
     a_i(t)=\alpha_i+\delta a_i(t)
\end{equation} 
Without loss of generality we set the population-averaged permanent and instantaneous susceptibility to 1: $\langle a_i(t) \rangle_i=\langle \alpha_i\rangle_i=1 $. Beyond its average value, the overall statistics of instantaneous  $\alpha(t)$ is properly defined only if that quantity is average over specified time window $\delta t$. Naturally,  its variation will gradually decrease as the  time widow increases. 

Individual reproductive number, $R_i$,  for  COVID-19 epidemic is (in)famously over-dispersed. This is a result of super-spreading, when a  majority of secondary infections are caused by a small fraction of index cases. The overdispersion reflects (i) variation of peak contagiousness level among individuals and (ii) dispersion of $a_i(t)$ which is effectively averaged over a timescale of the peak infection  period (approximately 2 days).

Importantly, according to Eq.(\ref{eq:Re}), the reproductive number  depends on correlations  of $a_i$ across a time scale of a single generation interval (on average, 4 to 5 days for COVID 19).  Thus, any variations in  $a_i(t)$ that do not persist over that timescale would be averaged out.
Here we introduce a simple model to account for temporal variation of social activity. In this model, $a_i$ may vary on short time scale,  relax to the persistent value for a given individual over certain relaxation  time, $\tau_s$:
\begin{equation}
     \dot{a_i}=\frac{\alpha_i-a_i}{\tau_s}+\eta_i(t)
\end{equation}
Here $\eta(t)$ is short-time noise that gives rise to time dependent fluctuations.  We set $\langle\eta_i(t)\eta_i(t')\rangle=\frac{2a_i(t)}{\tau_s k_0}\delta(t-t')$, which gives rise to individual diffusion in $a_i$ space with diffusion coefficient proportional to $a_i$. The  evolution of population with a given value of persistent activity $\alpha$ in that  space is given by  the following Fokker-Plank Equation:   
\begin{equation}
  \label{FP}
    \dot {\Psi}_\alpha (a,t)=   \frac{1}{k_0\tau_s}\frac{\partial^2\left(a \Psi(a,t)\right)}{\partial a^2} + \frac{1}{\tau_s}\frac{\partial \left((a-\alpha)\Psi_\alpha(a,t)\right)}{\partial a}
\end{equation}

The steady state solution  to this equation gives a  probability density function (pdf) for  $a$, which turns out to be a commonly used  gamma distribution:
\begin{equation}
\label{eq:Fokker}
  \Psi_{\alpha}(a,t)=f_\alpha(a) =\frac{a^{\alpha k_0-1}e^{-k_0 a}}{\alpha^{\alpha k_0}\Gamma(\alpha k_0)}
\end{equation}
Note that the  statistics of superspreading events is commonly modeled assuming the very same distribution for individual reproduction number, $R_i$.  This gives a strong empirical support to the chosen model, in particular to the choice of diffusion coefficient to be proportional to $\alpha$. It also allows us to partially calibrate the model. Reported dispersion parameter associated with  superspreading events for  COVID 19 is in the range of $0.1$ to $0.3$ \cite{Super_Kucharski,Vespigniani2020}. Note however that our parameter $k_0$ is expected to be larger than $k$, i.e. has a smaller dispersion. This is because variations of  $a(t)$ over the timescale shorter than a single generation interval would be averaged out according to Eq. (\ref{R0}),  while the superspreading statistics effectively probes it over a shorter time interval of the infectivity peak in a single individual. The latter could be further enhanced by a variation of individual contagiousness e.g. due to biological factors. 

It is well known that  the mean reproduction number $R_0$ in a heterogeneous population  depends on the second moment of distribution of $\alpha$ (in network epidemic  models it is related to individual degree). However, there is an important modification to that result for time-dependent $a(t)$:
  \begin{equation}
  \label{R0}
R_0=  \int_0^\infty \left\langle a_i(t) \beta_i(t+\tau)\right\rangle_i d\tau =R\langle \alpha_i^2 \rangle +\int_{0}^{\infty}  \left \langle C_i(\tau) \delta a_i(t)\delta a_i(t+\tau)  \right\rangle_{i}d \tau
 \end{equation} 
 Here  $R=\langle \int C_i(\tau)d\tau \rangle_i$ is the net infection transmission probability of an average person. This gives: 
 \begin{equation}
 R_0=\langle \alpha_i^2 \rangle +\mu k_0^{-1} 
 \end{equation}
 Here factor $\mu$ is related to the   Laplace transform of average contagiousness  profile, $K_0(\tau)=\langle C_i(\tau) \rangle_i/R$. 
 \begin{equation}
     \mu=\int_{0}^{\infty}  K_0(\tau) e^{-\tau/\tau_s}d \tau
 \end{equation} 
Note that, according to Eq.(\ref{eq:gen_int}),   generation interval pdf $K(\tau)$ is close, but not identical to $K_0(\tau)$:
\begin{equation}
K(\tau)=\left(1+\frac{e^{-\tau/\tau_s}-\mu}{k_0\langle \alpha_i^2 \rangle+\mu}\right)K_0(\tau)
 \end{equation} 
 Specifically, for the case of SIR model, i.e.  $K_0(\tau)\sim e^{-\tau/\tau_0}$, one obtains: $\mu=1/(1+\tau_0/\tau_s)\approx 1/(1+\tau_g/\tau_s) $.  Here   mean generation interval $\tau_g$ is given by
 \begin{equation}
 \tau_g=\frac{\tau_0(\langle \alpha_i^2 \rangle +\mu^2 k_0^{-1}) }{\langle \alpha_i^2 \rangle +\mu k_0^{-1} }\approx \tau_0\left(1-\frac{\tau_0}{\tau_s(1+k_0 {\langle \alpha_i^2 \rangle})}\right)
 \end{equation}
In this SIR case, one can assign  each  person a state variable $I_i$ set to  $1$ when the individual is infectious   and $0$ otherwise. This allows to  describe the  epidemic dynamics in terms of activity-weighted fraction of the infected population,  $I(t)=\langle I_ia_i(t)\rangle/\langle a_i^2\rangle$. Note that variable $j(t)$ and hence the strength of infection are proportional to it:
\begin{equation}\label{eq:J_I}
J(t)=M(t)j(t)=\frac{R_0M(t)I(t)}{\tau_g}
\end{equation}
    
\subsection{Mapping on quasi-homogeneous dynamic system}
Let $S_\alpha(a,t)$ be the fraction of  susceptibles among the sub-population with persistent activity level $\alpha$ and given value of $a$, at time $t$.  Change of function $S_\alpha(a,t)$ is driven by two effects: (i) depletion of susceptible population due to infection and (ii) diffusion of individual in $\alpha$ space. By substituting $\Phi_\alpha(a,t)=f_\alpha(a) S_\alpha(t)$ into Fokker-Plank Eq. (\ref{FP}), and adding the infection term with rate  $-a(t)$, we obtain an evolution equation for $S_\alpha$; 
\begin{equation}
\label{S_master}
    \dot {S}_\alpha(a,t)=-a S_\alpha(a,t) J(t) + \frac{a}{k_0\tau_s}  \frac{\partial^2 S_\alpha(a,t)}{\partial a^2} + \left(\frac{\alpha-a}{\tau_s}\right)\frac{\partial S_\alpha(a,t)}{\partial a}
\end{equation}

This equation can be solved by using the following ansatz:
\begin{equation}
S_\alpha(a,t)=\exp\left[-Z(t)\alpha- k_0h(t)a\right] 
\label{eq:s_alpha_anzats}
\end{equation}
Here $Z(t)$ is a measure of persistent heterogeneity: the larger it is, the more is the difference in depletion of susceptible among subpopulations with different $\alpha$'s, i.e. various average levels of social activity. On the other hand,  $h(t)$ parameterizes the transient heterogeneity within each of these  subpopulations.    In the long run, this type of heterogeneity disappears due to evolution in $a$-space, thus $h(t)$ asymptotically approaches $0$ as $t \to \infty$. 
Substituting Eq. (\ref{eq:s_alpha_anzats}) into Eq. (\ref{S_master}) results in simple equations for both $Z(t)$ and $h(t)$:
\begin{align}
\label{dzdt}
\dot{h}&=\frac{J(t)}{k_0}-\frac{h(t)(1+h(t))}{\tau_s}  \\
\label{S_star}
\dot{Z}&=\frac{ k_0 h(t)}{\tau_s} 
\end{align}

The renewal equation Eq. (\ref{attack0}) for $j(t)$ completes our quasi-homogeneous description of the epidemic dynamics. However, to fully close this system of equations, one needs to express the effective reproduction number,    $R_e$, in terms of the functions $M(t)$, $Z(t)$ and $h(t)$. This is done by substituting  the ansatz, Eq. (\ref{eq:s_alpha_anzats}), into Eq. (\ref{eq:Re}). We perform this calculation in two steps, by first finding the effective  number $R_\alpha$ for a sub-population with average level of activity $\alpha$, followed by averaging over persistent heterogeneity. This gives
\begin{equation}
  R_\alpha=\int_0^\infty a (\alpha+\mu(a-\alpha))f_\alpha(a) e^{-Z(t)\alpha- k_0 h(t)a}d a 
= \frac{\alpha R\left(\alpha+\mu k_0^{-1}+h(1-\mu)\right)e^{-\tilde{Z}\alpha}}{\left(1+h(t)\right)^2}
\end{equation}
Here  
\begin{equation}
\tilde{Z}=Z+k_0\ln(1+h)
\end{equation}
Note that 
\begin{equation}
\label{tildeZ}
\dot{\tilde{Z}}=\frac{J(t)}{1+h(t)}
\end{equation}
The averaging over persistent heterogeneity, under the assumption that $\alpha$ obeys the gamma distribution,  $p(\alpha)\sim \alpha^{\kappa-1}e^{-\kappa \alpha}$, yields
\begin{equation}
\label{eq:Re_pers}
  R_e(t)=M(t)\int_0^\infty R_\alpha  p(\alpha) d\alpha = \frac{\chi+\left(1-\chi)(1+k_0h(\mu^{-1}-1)\right)\left(1+\kappa^{-1}\tilde{Z}(t) \right)R_0M(t)}{\left(1+\kappa^{-1}\tilde{Z}(t) \right)^{2+\kappa}\left(1+h(t)\right)^2}
\end{equation}
Here
\begin{equation}
   \chi=\frac{1+\kappa^{-1}}{1+\kappa^{-1}+\mu k_0^{-1}} 
\end{equation}

Similarly, we calculate $S$, which ends up having the same form as in the  model with  persistent heterogeneity \cite{hetero_PNAS}:  
\begin{equation}
\label{eq:S_pers}
  S(t)=\int_0^\infty\int_0^\infty p(\alpha) f_\alpha(a) e^{-Z(t)\alpha- k_0 h(t)a}da d\alpha =\frac{1}{\left(1+\kappa^{-1}\tilde{Z}(t) \right)^\kappa}
\end{equation}

By comparing Eqs.(\ref{eq:Re_pers}) and (\ref{eq:S_pers}) we obtain $R_e$ in terms of  $S$ and $h$:
\begin{equation}
\label{eq:Re_vs_S}
R_e(t)= \frac{R_0M(t)S^{\lambda}q_\chi(S,h)}{\left(1+h(t)\right)^2}
\end{equation}
Here 
\begin{align}
    q_\chi(S,h)&=(1-\chi)\left(1+k_0h(\mu^{-1}-1)\right)S^{-\chi/\kappa}+\chi S^{(1-\chi)/\kappa}\approx 1\\
    \lambda&=1+\frac{1+\chi}{\kappa}=\frac{\left(1+\kappa^{-1}\right)\left(1+\mu k_0^{-1}+2\kappa^{-1}\right)}{1+\mu k_0^{-1}+\kappa^{-1}}
\end{align}
Note that  for most practical purposes, one can set $q_\chi(S,h)=1$.     The parameter  $\lambda$ is  the \lq\lq immunity factor" that  emerged in the context of our earlier  study of persistent  heterogeneity \cite{hetero_PNAS}. In that case, $\lambda=1+2/\kappa$ appears as a scaling exponent in relationship between effective reproduction number  $R_e(t)$ and the fraction of the  susceptible population $S(t)$: $R_e=R_0 M S^\lambda$. Eq. (\ref{eq:Re_vs_S})  generalizes that result. 

Eqs.(\ref{attack0}), (\ref{dzdt}), \ref{eq:Re_vs_S}, (\ref{tildeZ})  give the full description of the epidemic dynamics in heterogeneous system.  For a particular case  of  SIR model ($K(\tau)\sim e^{-\tau/\tau_g}$), we obtain a  3D dynamical system, in terms of variables $I(t)$, $S(t)$ and $h(t)$:
\begin{align}
     \label{eq:JSH1}
     \frac{dI}{dt}&=  \frac{JS^{\lambda}}{\left(1+h\right)^2}- \frac{I}{\tau_g}\\
    \label{eq:JSH2} \frac{dS}{dt}&= -\frac{JS^{1+1/\kappa}}{(1+h)} \\ 
    \label{eq:JSH3} \frac{dh}{dt}&=\frac{J}{k_0}-\frac{h(1+h)}{\tau_s}
\end{align}
Here, infection strength $J(t)$ is proportional to the activity-weighted fraction of susceptible population $I(t)$, and the mitigation profile $M(t)$, as given by Eq. (\ref{eq:J_I}).   
Eq. (\ref{eq:JSH2}) was derived by combining  Eq. (\ref{tildeZ}) and Eq. (\ref{eq:S_pers}). Alternatively, after  substituting result of integration of    Eq. (\ref{tildeZ}) into (\ref{eq:S_pers}), one gets the explicit formula for $S(t)$:
\begin{equation}
S(t)=\left(1+\kappa^{-1}\int_{-\infty}^t\frac{J(t')dt'}{1+h(t')}\right)^{-\kappa}
\end{equation}

\subsection{Waves and plateaus}
According to Eq. (\ref{eq:JSH1}),  the  combined driving force of the epidemic is $R^*=R_0M(t)S^\lambda(t)$. It includes both the effects of mitigation $M(t)$ and  suppression associated with the build up of the long-term herd immunity.  First,  we assume $R^*$ to be fixed or change very slowly (adiabatically), i.e. on the timescales longer than $\tau_s$. In that case,    $J(t)$ and  $h(t)$ trail  the driving force $R^* (t)$, staying close to the corresponding adiabatic fixed point $(J^*,h^*)$ in  their 2D phase space:
\begin{align}
    h^*&=\sqrt{R^*}-1\\
    J^*&=\frac{k_0h^*(1+h^*)}{\tau_s}
\end{align}
The stability of this adiabatic fixed point, and the more  rapid epidemic dynamics can be described by linearizing    
Eqs. (\ref{eq:JSH1}, \ref{eq:JSH3}) and (\ref{dzdt}) around $(J^*,h^*)$,  i.e. by assuming   $h(t)=h^* + \delta h(t)$ and $J(t)=J^*+\delta J(t)$:
\begin{equation}
      \frac{d}{dt}\begin{pmatrix}
    \delta h \\
     \delta J
    \end{pmatrix}= \frac{1}{\tau_s}
  \begin{pmatrix}
    -(1+2h^*) & \tau_s/k_0\\
  -2k_0\gamma h^*& 0
    \end{pmatrix}\begin{pmatrix}
     \delta h\\
    \delta J
    \end{pmatrix}
\end{equation}
The eigenmodes of this linearized system are both stable, but the rates have  substantial imaginary components:   
\begin{equation}
    r_{\pm}= -\frac{1+2h^*}{2\tau_s}\pm i \sqrt{\frac{2\gamma h^*}{\tau_s}-\frac{(1+2h^*)^2}{4\tau_s^2}}
\end{equation}
This indicates that relaxation towards  point  $(J^*,h^*)$ has a pronounced oscillatory character. The period of the  oscillations is  
\begin{equation}
T\approx \pi\sqrt{\frac{2 \tau_s}{\gamma h^*}}\approx \pi\sqrt{\frac{2 \tau_s}{\gamma (\sqrt{R^*}-1)}}
\end{equation}
The amplitude of the oscillations decays  with the time constant $2\tau_s/(1+2h^*)$. This oscillatory behavior would manifest itself as multiple epidemic waves. In reality, the dynamics is more complicated since rapid changes of $M(t)$, e.g. due to seasonal effects, government and societal response to the epidemic, would additionally modulate it.   

The assumption of $R^*=R_0M(t)S^\lambda(t)$ being fixed is not, of course,   realistic. In particular, the mitigation factor  $M(t)$  may have both slow and fast variations.  On top of that,  the  dependence of $R^*$ on   $S(t)$  creates a negative feedback  suppressing  the forcing on the long run. For a constant mitigation $M$, there is a line of  fixed points $(J,S,h)=(0,S,0)$, for any  $S\le S_{0}=\left(R_0M\right)^{-1/\lambda}$.  Here $1-S_0$ represents  the long-term herd immunity threshold (HIT) for a given  mitigation level $M$.  There is one particular  solution $(\tilde{J}(t), \tilde{S}(t), \tilde{h}(t))$ corresponding  to all three variables slowly evolving  in  such a way that $R_e$ stays close to   $1$  at all  times, eventually reaching the HIT point, $(0,S_0)$. As follows from the above  stability analysis, this solution acts as an attractor, with any  trajectory in   $(J, S, h)$ space converging towards it, unless perturbed by variations in mitigation $M(t)$. To construct that solution, we 
%
%
%
set the growth rate for $I(t)$ in   Eq.(\ref{eq:JSH1}) to $0$, and use Eqs (\ref{eq:JSH2})-(\ref{eq:JSH3}) to calculate the corresponding evolution of  $h(t)$:  
\begin{align}
\left(\tau_s+\frac{\tilde{\tau}-\tau_s}{(1+\tilde{h}(t))^{2\nu}}\right)\dot{\tilde{h}}&=\tilde{h}(t)(1+\tilde{h}(t))\\
    \tilde{h}(t)&\approx\frac{1}{(1+2\nu)(e^{(t-t_0)/\tilde{\tau}}-1)}
\end{align}
Here $\nu=\frac{1}{\lambda\kappa}$, and    
\begin{equation}
    \tilde{\tau}=\tau_s\left(1+\frac{2\nu(R_0M)^{\nu}}{k_0}\right)\simeq \frac{\tau_s}{k_0}
\end{equation}

Remarkably, under assumption of strong overdispersion, $k_0 \ll 1$,  the  emergent timescale $\tilde{\tau}$ is   significantly longer than the social rewiring time, $\tau_s$. This long timescale corresponds to a slow process of individuals trapped in the low activity state, $a(t)\le k_0$,  transitioning to the high activity level $a \ge 1$.  Respective evolution of ${\tilde S}(t)$ and ${\tilde J}(t)$ are given by: 
\begin{align}
  \tilde{S}(t)&= S_0\left(1+\tilde{h}(t)\right)^{2/\lambda}\\
\tilde{J}(t)&\approx \left(\frac{1}{\tau_s}-\frac{1}{\tilde{\tau}}\right)k_0\tilde{h}(t)(1+\tilde{h}(t)) 
\end{align}

%
%
%

\subsection{Waning of biological immunity}
Our equations could be easily modified to account for the waning of biological immunity. This adds a new term in Eq. (\ref{eq:JSH2}) which becomes:
\begin{equation}
\label{eq:JSH2_mod}
\frac{dS}{dt}=-\frac{JS^{1+1/\kappa}}{(1+h)} +\frac{1-S}{\tau_b}
\end{equation}
Here $\tau_b$ is the lifetime of biological immunity,  which we set to $5$yrs throughout this work. 
The last term $\frac{1}{\tau_b}(1-S)$ describes the rate at which the recovered population (fraction $1-S$) reverts back to the susceptible state. The endemic steady state can be found by setting time derivatives Eqs. (\ref{eq:JSH1}),(\ref{eq:JSH3}) and (\ref{eq:JSH2_mod}),  to $0$. Under the assumption that $\tau_b\gg \tilde{tau}$, the endemic point in  $(S,J,h)$ is given by 
\begin{align}
    J_{en}& \approx \frac{1-S_{HIT}}{\tau_bS_{HIT}^{1+1/\kappa}}\\
    h_{en}& \approx \tilde{\tau}J_{en}=\frac{\tilde{\tau}}{\tau_b}\frac{1-S_{HIT}}{S_{HIT}^{1+1/\kappa}}\\
    S_{en}& = S_{HIT}(1+h_{en})^{2/\lambda}\approx S_{HIT}
\end{align}
Here $S_{HIT}={R_0M}^{1/\lambda}$ corresponds to the HIT.
\subsection{Seasonal forcing}
Seasonal effects are commonly described as simple $\sin$-shaped modulation of reproductive number \cite{Neher2020}. In this work, we used a combination of sigmoidal functions to model transition between \lq\lq winter" and \lq\lq summer" values of $M(t)$: 
\begin{equation}
    M_{s}(t)=1+\sigma\sum_{n=0}^\infty\left[ 1-\tanh \left(\frac{t-t_{\rm spring}+nT} {\Delta}\right)+\tanh \left(\frac{t-t_{\rm fall}+nT}{\Delta}\right)\right]
\end{equation}
Here $T= 1$yr, time parameters  $t_{\rm spring}<t_{\rm fall}$  and $\Delta$ determine the timing of and sharpness of  winter-summer-winter transitions. $\sigma$ determines the amplitude of seasonal changes. In particular, $\sigma=0.25$ in Fig.\ref{fig:diagram}, and ranges between $0.25$ and $0.35$ in our fits of epidemic dynamics for different US regions, Fig. \ref{fig:southwest}.

\subsection{Implementation of the  agent-based model}
All simulations for the agent-based model use 1 million agents and 3 simulation replicates. For each agent in the simulation, at each time step, the social activity follows the stochastic dynamics described in Eq. \ref{Langevin}. After that, the overall force of infection is computed using
\begin{equation}
    J(t) = \frac{R_0 M(t)}{\tau_g\langle a(t)^2 \rangle_i} \frac{1}{N} \sum_i a_i I_{i} \qquad ,
\end{equation}
where $I_{i}$ is binary and used to denote whether or not the agent is infectious, $N$ is the number of agents in the simulation. For a susceptible agent $i$, the chance of being infected in one simulation step is $a_i(t) J(t) dt$ which is proportional to the force of infection, his/her activity $a_i(t)$, and $dt$ - the length of the time step used in our simulations. For an infectious agent, the probability of recovering from the infectious state in one simulation step is $\gamma dt$. When the waning of biological immunity is ignored, recovered agents will always stay in the recovered state and cannot be infected again.

\begin{table}[ht!]
\caption{
    Set of fixed model parameters
}
\begin{tabular}{p{1 cm}p{1 cm}p{1.5 cm}p{1.5 cm}p{1 cm}}
\hline
\hline
$k_0$ & $\kappa$ & $\tau_g$ & $\tau_s$ & $\tau_b$                \\
\hline
$0.4$ &$2$ & $5$ days & $30$ days & $5$ yrs                   \\
\hline\hline
\end{tabular}
\label{table:parameters}
\end{table}

\end{document}